\documentstyle[]{aipproc}

\newcommand{\be}{\begin{equation}} \newcommand{\ee}{\end{equation}}
\newcommand{\bea}{\begin{eqnarray}} \newcommand{\eea}{\end{eqnarray}}

\newcommand{\Uno}{{\rm 1\hspace{-0.9mm}l\hspace{-1.0mm}}}

 \input epsf 

\begin{document}
\title{On Global Analysis of Duality Maps}

\author{I. Mart\'{\i}n \thanks{ Invited talk at I-SILAFAE in a splendid site of Yucat\'an.}
and A. Restuccia \thanks{Plenary talk at I-SILAFAE, Merida,Mexico.}}
\address{Universidad Simon Bolivar, Departamento de Fisica,\\ Caracas, 
Venezuela. e-mail: isbeliam@usb.ve, arestu@usb.ve}
\maketitle
\begin{abstract}
A global analysis of duality transformations is presented. It is shown that 
duality between quantum field theories exists only when the geometrical structure 
of the quantum configuration spaces of the theories comply with certain precise conditions.
Applications to S-dual actions and to T duality of string theories and D-branes 
are briefly discussed.  It is shown that a new topological term in the dual open
string actions is required. We also study an extension of the procedure to
construct duality maps among abelian gauge theories to the non abelian case. 
\end{abstract}
  
Duality transformations were introduced by Dirac and extended later on by 
Montonen and Olive.  More recently were used by Seiberg and Witten \cite {1} to 
relate the weak and strong coupling regime in the analysis of the low 
energy effective action of the N=2 SUSY SU(2) Yang-Mills. This approach to 
non-perturbative QFT was then introduced in string theory with spectacular 
success.  It has been shown that the strong coupling regime of one string 
theory can be mapped to the weak coupling regime of another perturbatively 
different string theory, giving rise to a possible unification of all 
string theories in the context of a hypothetical M-theory.
In this lecture, we analyse the duality transformations from a global point 
of view.  This approach requires the introduction of a more general 
geometrical structure than the associated to line bundles over a general 
euclidean base manifold.  We describe \cite {2} the general structure of 
higher order line bundles, and define over them dual maps between theories 
described locally by p-forms.

Duality maps for theories with p-forms have been discussed in \cite {3} 
and appear naturally in the description of D-brane theories.  However,
the global aspects of the configuration space of these local p-forms was
never described. We show that the local analysis used in \cite {3} is not 
enough to ensure quantum duality equivalence and give the necessary 
conditions to achieve it.  The interesting result related to this global 
structure is that duality between theories
of local p-forms and d-p-forms not only imply the quantization of couplings,
the known generalized Dirac quantization condition, but also determine completely, from a global point 
of view, the configuration space of these local p-forms.  These spaces 
are defined in terms of local p-forms with non trivial transitions on higher 
order bundles.
In the first part of the lecture we explain the global approach for the 
Maxwell theory formulated over a general base manifold.  We then give the 
general results concerning the higher order bundles and discuss some 
applications to D-brane theories.  To do so we first consider the duality 
analysis for open bosonic strings.  We prove that the dual open string 
action requires a new topological term in order to obtain the correct dual 
boundary  conditions.
In the second part of the lecture, we extend these ideas to the case of non abelian 
dualities.
\section{Electromagnetic Duality}
The action of Maxwell theory over a 4-dim base manifold X, compact 
euclidean and orientable, is
\be
I(F(A))=\frac{1}{8\pi}\int_{X}d^4x\sqrt{g}[\frac{4\pi}{e^2}F^{mn}F_{mn}+i\frac{\theta}{4\pi}
\frac{1}{2}\epsilon_{mnpq}F^{mn}F^{pq}]
\ee
where F is the curvature of a 1-form connection A on a U(1) bundle.  This 
action may be rewritten in terms of the complex coupling $\tau= 
\frac{\theta}{2\pi}+\frac{4i\pi}{e^{2}} $

\be I_\tau(A)=\frac{i}{8\pi}\int_X{}d^4x\sqrt{g}[\bar{\tau}F_{mn}^+F^{+mn}
-\tau{}F_{mn}^-F^{-mn}]\ ,
\ee
In order to construct the dual map we introduce an equivalent formulation to 
(1) or (2).  We consider the action

\be 
I(\Omega{})=\frac{i}{4\pi}[\bar{\tau}(\Omega^+,\Omega^+)-\tau{}(\Omega^-,\Omega^-)]
\ee
where $\Omega$ is a global 2-form satisfying the constraints
\bea 
d\Omega{}=0\label{eq:eq4}\\
 \oint_{\Sigma_2^{I}}\Omega=2\pi n^{I}
\eea

The motivation to introduce the global constraint (5) is that by Weil's 
theorem  constraints (4) and (5) ensure the existence of a unique 
complex line bundle and a connection on it -not necessarily unique- whose 
curvature is $\Omega$.   (3), (4) and (5) represent then an equivalent formulation 
to (1) in terms of the configuration space of 2-forms.  It is relevant to determine how many connections giving $\Omega$ 
can be constructed for a given line bundle. In this case it is given by 
$H^{1}(X,R)/H^{1}(X,Z) $.  The cohomology classes take into account all canonical gauge 
equivalent connections, while $H^{1}(X,Z)$ counts for the equivalence under "large" gauge 
transformations.

Having determined the exact correspondence between connections over line 
bundles and global 2-forms constrained by (4) and (5), we have to 
introduce now the correct Lagrange multiplier.  It must have also a 
precise global structure in order to account for the global constraint (5). 
It can be shown \cite{2} that it may be expressed in terms of a 1-form 
connection V over the dual line bundle, provided summation over all dual 
bundles and all gauge inequivalent connections over every line bundle is 
performed in the functional integral.
 The resulting functional integral in 
terms of A must also be an integral on all line bundles and all gauge 
inequivalent connections on every line bundle.  The important point to 
emphasize here is that the configuration spaces for the 1-form connections A 
and V are uniquely determined from the duality equivalence.  In this sense 
the requirement to theory (1) of having a dual formulation determines 
completely the global structure of its configuration space.  The 
quantization of magnetic charge is then only one consequence of this 
global structure.
The resulting action after the introduction of the Lagrange multiplier is 
given by
\be
{\cal{I}}(\Omega,V)=I(\Omega)+\frac{i}{2\pi}\int_{X}W(V)\wedge\Omega 
\ee.

From (6) one may integrate on V and regain (3), (4) and (5) and after 
solving the constraints (4) and (5) one obtains (1). We can also integrate on $\Omega$ and obtain 
the dual action in terms of V.   The partition function of both quantum equivalent formulations is 
obtained in the standard way, with the known \cite{3} result  
\be {\cal{Z}}(\tau{})={\cal N}
\tau^{-\frac{1}{2}B^-_2}\bar{\tau}^{-\frac{1}{2}B^+_2}{\cal{Z}}(-\frac{1}{\tau}) \label{eq:dualz}
\ee
\noindent{}where $B^+_k$ and $B^-_k$ are the dimensions of the spaces of selfdual and antiselfdual
$k$ forms.
\section{Duality on higher order bundles}
  
The generalization of the above construction may be analysed \cite {2} by considering a globally 
defined p-form over X satisfying
\begin{eqnarray} 
dL_{p} & =& 0 \nonumber\\ 
\oint_{\Sigma_{p}^{I}}L_{p} &=& 
2\pi n^{I}. 
\end{eqnarray} 
Let us consider $p=3$. We take an open covering of $ X : \{U_{i}, i\in I\}$. Without 
loosing generality we may consider every open set and its intersections to 
be contractible to a point .  On $U_i$ we have

\begin{equation}
 L_{3}=dB_{i} 
\end{equation}

and on $U_i\cap U_j\neq\phi$ 

\begin{eqnarray} 
dB_{i} &= &dB_{j}  \nonumber\\ B_{i}& = & 
B_{j}+ d\eta_{ij}
\end{eqnarray}
where $B_i$ is a 2-form with transitions given by (10), $\eta_{ij}¥$ being a 
local 1-form defined on $U_i\cap U_j\neq\phi$. On $U_i\cap U_j\cap U_k$ we 
obtain 
\begin{eqnarray}
 L_{1} &\equiv & \eta_{ij}+\eta_{jk}+\eta_{ki}\nonumber \\
  d L_{1}&  = &0 
\end{eqnarray}

From (8) we have
\begin{equation} 
\int_{\Sigma_{1}}L_{1}=2\pi n,
\end{equation} 

where $\Sigma_1$ is a close curve on $U_i\cap U_j\cap U_k$.  From (11) and 
(12) we obtain  a 1-form $L_1$ defined over $U_i\cap U_j\cap U_k$ 
satisfying 
\begin{eqnarray} 
dL_{1} & =& 0\\
\int_{\Sigma_{1}}L_{1} &=& 2\pi n
\end{eqnarray}
which yields an uniform map from 
\begin{equation}
 U_i\cap U_j\cap U_k \rightarrow U(1)
\end{equation}

The interesting property not present in the previous discussion is that 
the 1-cochain is now defined as
\be g:(i,j) \rightarrow g_{ij}(P,{\cal C}) \equiv \exp 
i\int_{\cal C}\eta_{ij}
\ee
where $\cal C$ is an open curve with end points $O$ (a reference point) 
and $P$.

 $g$ associates to$(i,j)$ a map $g_{ij}(P,\cal C)$ from the path space over $U_i\cap U_j$ to 
the structure group
U(1).

Notice that the 1-form $\eta_{ij}$ cannot be integrated out to obtain a 
transition function as in
the case of a line bundle. However, we have
\be 
\delta{}g_{ijk}=g_{ij}g_{jk}g_{ki}=\exp
i\int_{O}^{P}L_{1} 
\ee
 which is precisely the uniform map $M$ previously 
defined in (15). (16) explicitly shows that the geometrical structure we are dealing with is not 
that of an usual $U(1)$bundle since the cocycle condition on the intersection of three open sets 
of the covering is not
satisfied. Starting from transitions functions $g_{ij}$ defined on the 
space of paths over
$U_i\cap U_j$, and acting with the coboundary operator $\delta$ we obtain 
the 2-cochain (17)
which is properly defined in the sense of $\breve{C}ech$. We may go further and 
consider in the
intersection of four open sets the action of the coboundary operator 
$\delta$ on 2-cochains.We obtain on $U_i\cap U_j\cap U_k\cap U_l$ a 3-cocycle condition in the 
sense of $\breve{C}ech$: 
\be 
\delta{}g_{ijkl}=g_{ijk}g^{-1}_{ijl}g_{ikl}g^{-1}_{jkl}=\Uno 
\ee

The construction leads then to local p-forms with non-trivial transitions 
defined by the higher order bundle.  Having extended the geometrical 
structure of line bundles we may then formulate over them duality maps 
generalizing the electromagnetic duality.

The action for the local p-form $A_p$  defined over open sets of a 
covering of X and with transitions defined over a higher order bundle is 
\be 
{S(A_p)}= \frac{1}{2}\int_X{F_{p+1}\wedge*F_{p+1}}  
\ee 

where $F_{p+1}$ is the curvature of $A_p$ and the coupling constant has 
been reabsorbed in $A_p$.

Let us now consider its dual formulation.  We introduce now the globally 
defined (p+1)-form  $L_{p+1}$ satisfying

\begin{eqnarray}
 dL_{p+1} & =& 0 \\
 \oint_{\Sigma_{p}^{I}}L_{p+1} &=& \frac{2\pi n^{I}}{g_p}.
\end{eqnarray} 
with action
\be {S}= \frac{1}{2}\int_X{}L_{p+1}\wedge 
*L_{p+1}{} 
\ee 
(20) and (21) ensure the existence of a bundle of order p+1 and a local p-form 
on it whose curvature is $L_{p+1} $.

The off-shell Lagrange problem of the above constrained system may be given 
by the action
\be
S(L_{p+1},V_{d-p-2}) = S(L_{p+1})+ i \int_X{}L_{p+1}\wedge W_{d-p-1}(V){} 
\ee
where $V_{d-p-2}$ is a local (d-p-2)-form with transitions over a higher 
order bundle satisfying a (d-p-1)-cocycle condition and with coupling 
$g_{d-p-2}$.

Funtional integration on $L_{p+1}$ yields
\be
 *L_{p+1} = - i W_{d-p-1} 
\ee
where $W_{d-p-1}$ is the curvature of $V_{d-p-2}$, and the dual action
 \be 
 S(V_{d-p-2}) =\frac{1}{2}\int_X{}W_{d-p-1}(V)\wedge *W_{d-p-1}{} 
\ee

The quantum equivalence between the two dual actions follows once we 
integrate on all bundles of order p generalizing the electromagnetic 
duality previously shown.  The quantization of charges arises directly from 
the global constraints needed for having a globally well defined higher 
order bundle. The configuration space of the local p-forms $A_p$ and its 
dual are globally determined, they are defined  over higher order bundles 
with cocycle condition of order $p+1$ which are classified by the integer 
numbers $n^I$ associated to a basis of integer  homology $\Sigma_I$ on X.  
For a given bundle of  order p+1 the different local antisymmetric fields 
up to gauge transformations are given by 
${{H^p(X,U(1))}/{H^p(X,Z)}}$.

These local antisymmetric fields with non trivial transitions appear 
naturally in the description of D-branes.  For example it has been \cite {4} 
conjectured that the d=11 5-brane action is given by
\begin{equation}
 S=-\frac{1}{2}\int_{X} d^{6}\xi \sqrt{-\gamma}[\gamma^{ij}\partial_{i}x^{M}
 \partial_{j}x^{N}\eta_{MN}+\frac{1}{2}\gamma^{il}\gamma^{jm}\gamma^{kn}F_{ijk}F_{lmn}-4]
\end{equation}
where $F=dA$ is the self dual 3 form field strength of a local 2-form 
potential A which has to be defined over a bundle of order 3 if non trivial 
topological effects are expected. It would be interesting to determine 
completely from a geometrical point of view the moduli space of the self 
dual potentials over this higher order bundle.  This problem is under 
study.

It is interesting to notice that dealing with D-brane theories, there are 
two different duality transformations involved.  One is obtained by 
following the approach we have described previously with respect to the 
local 2-form A in (26).  Because of the self duality condition, the 
curvature $F_3$ may be identified with $W_3$.  The other duality arises by 
following the same approach but with a different interpretation for the 
global constraint, it is now related to the compactification condition on 
some of the coordinates on the target space.  To show it in some detail we 
explain the duality transformation on the worldsheet of the string 
theory, and finally comment on the D=11 supermembrane, D=10 IIA Dirichlet 
supermembrane duality transformation which involve a compactification of 
$\chi^{11}$, say, on $S^1$ and nontrivial line bundles over the worldvolume from 
the other side.  That is, both kind of global constraints appears in the 
duality map.
\section{T Duality}
We discuss now the duality maps between first quantized string theories 
emphasizing the global constraint in the construction.

The string action is
\begin{equation}
S(\chi)	=\frac{1}{2\alpha^{\prime}}\int_{\Sigma}d^{2}\xi\sqrt{g}g^{ij}\partial_{i}\chi^{\mu}
	\partial_{j}\chi_{\mu}
\end{equation}
where $g^{ij}$ is the world sheet metric and $\xi^i$ , i=1,2 are the local 
coordinates of the Riemann surface $\Sigma$ of a fixed topology.  
We analyse first the closed string theory with one coordinate $\chi$ 
compactified over $S^1$.  Associated to that coordinate we introduce a 
contrained 1-form $L=L_id\xi^i$ satisfying
 \begin{eqnarray}
 dL & = & 0 \\ 
 \oint_{{\cal C}^{I}}L & = & 2\pi n^{I}R 
\end{eqnarray} 
where ${\cal C}^I$ denotes a basis of the integer homology of dimension 1 over the 
worldsheet.  Contraint (28) implies L is a closed 1-form, while (29) ensures 
the compactification over $S^1$, R is the compactification radius.  The 
solution to (28) and (29) is the string map $\chi(\xi^1,\xi^2)$.
We introduce Lagrange multipliers associated to constraints (28) and (29) and 
obtain the quantum equivalent action
\begin{equation}
	S(L,V)=\frac{1}{2\alpha^{\prime}}\int_{\Sigma}L\wedge ^{\ast}L+\frac{i}{\alpha^{\prime}}
	\int_{\Sigma}L\wedge  W(V)
\end{equation}

where $V(\xi^1,\xi^2)$ is the dual map to $\chi (\xi^1,\xi^2)$:

\begin{eqnarray}
	W(V) & = & dV\\
	\oint_{{\cal C}^{J}} & = & 2\pi m^J R^{\prime}
\end{eqnarray}

(32) is uniquely determined to obtain quantum equivalence between $S(L,V)$ 
and $S(\chi)$.

Following the same arguments as in the S duality approach  we obtain, in order to recover (29) 
from (30), after summation on all n in  the functional integral,
\begin{equation}	
R^\prime=\frac{\alpha ^ \prime}{R}
\end{equation}
 
That is the dual radius arises directly from the off-shell construction of the 
dual action.
From (30) we obtain the standard on-shell duality relation

\begin{equation}	
^\ast L + iW=0
\end{equation}
 From (30) after functional integration on L we obtain
\begin{equation}
	S(V)=\frac{1}{2\alpha^{\prime}}\int_{\Sigma}W(V)\wedge ^{\ast}W(V).
\end{equation}

The duality between $L=d\chi$ and $W=dV$ is resumed in the global constraints 
(29), (32) and (34).  Notice that (32) is uniquely determined from the 
off-shell construction while (29) implies the compactification of $\chi$ on 
$S^1$ of radius R.
The quantum equivalence between (27) and (35) has been shown for any compact 
Riemann surface $\Sigma$ hence the T-duality is valid order by order in the 
perturbative expansion of closed string theories.
We now discuss the duality of open string theories.
The standard open string boundary condition arises from (27) by considering 
the stationary points of $S(\chi)$.  Its variation yields a boundary term
\begin{equation}
\displaystyle \left. \left (\delta \chi \partial_{i}\chi n^{i} \right) 
\right|_{ \partial \Sigma }
\end{equation}
It can be annihilated by assuming
\begin{equation}
	\partial_{i}\chi^{\mu}n^{i}=0 
\end{equation}

This boundary condition together with the usual string field equation gives 
an stationary point of (27) with respect to the space of variations 
$\delta \chi$ which are arbitrary even on the boundary.
If instead we consider the space of maps $\chi$ restricted by a boundary 
condition and look for a stationary point of (27) restricted to that space, 
then 
\begin{equation}
\displaystyle \left.	\chi^{\mu}\right|_{ \partial \Sigma }=cte
\end{equation}
would be also a solution, since  then $\delta \chi =0$.  In this case one can 
have even a mixture of Dirichlet and Newmann conditions on the boundary as 
an acceptable solution.
We will discuss the construction of the dual string action on the first 
case and show that a topological action term has to be added to (35) in 
order to have a dual action whose stationary points yields the dual 
boundary condition to (37).  
Notice that from the duality relation (34) one obtains
\begin{equation}
n\cdot L=0 \rightarrow t\cdot W=0
\end{equation}
where t is tangent to the boundary.  However from (35) if we consider 
arbitrary variations on the boundary we get
\begin{equation}
n\cdot W=0
\end{equation}

We thus must modify (35) and consequently (30).
We consider
\begin{equation}
\tilde{S}(L,V,Y)=S(L,V) + \frac{i}{\alpha^{\prime}}\int_{ \Sigma }F(Y)\wedge W(V)
\end{equation}
where $F=dY$  and Y is a map onto $S^1$.  The new term in the action is a 
pure topological one.  It does not modify the field equations, only 
contributes to the boundary terms.  All the local dependence of 
$Y(\sigma)$ can be gauged away, only the boundary contribution remains.
The boundary terms in the variation of (41) are
\begin{eqnarray}
	\displaystyle \left	(\delta V (L+F)\right|_{\partial \Sigma } & = & 0\\
{	\displaystyle \left	(\delta Y W)\right|_{\partial \Sigma }} & = & 0 
\end{eqnarray}
which imply V=cte over any connected part of the boundary, and
\begin{equation}
\displaystyle \left	(L+F(Y))\cdot t \right|_{\partial\Sigma }	
\end{equation}
  
(44) does not add any restriction to L. It only determines $F(Y)$ on 
$\partial \Sigma $.
After integration on L we obtain
\begin{equation}
\tilde{S}(V,Y)= \frac{1}{2\alpha^{\prime}}\int_{\Sigma }W(V) \wedge ^\ast W(V) +
\frac{i}{\alpha^{\prime}}\int_{ \Sigma }F(Y)\wedge W(V)
\end{equation}

We will consider now that V and $\chi$ are maps onto $S^1$ with 
compactification radius R' and R respectively.  This implies that V=C on 
the $\sigma =0$ boundary and $V=C+2\pi nR^{\prime}$ in the $\sigma=\pi$ boundary.  We 
will show quantum equivalence between (27) with boundary condition 
$d\chi\cdot n=0$ 
and (45) with boundary condition $dV\cdot t=0$.
Starting from (41) integration on V yields (27)  and we are left with the 
boundary terms
\begin{equation}
\frac{i}{\alpha^{\prime}}C \int_{\partial \Sigma }[L+F(Y)] + 
\frac{i}{\alpha^{\prime}}2 \pi nR^\prime \int_{\sigma = \pi}[L + F(Y)]
\end{equation}
integration on C and summation on n yield
\begin{equation}
\delta \left( \int_{\partial \Sigma }[L+F(Y)] \right) \sum_{m} \delta \left( 
\frac{R^\prime }{\alpha^\prime} \int_{\sigma = \pi}[L + F(Y)] + 2 \pi m \right)
\end{equation}

They imply that
\begin{equation}
\int_{\sigma = \pi}L= \left[-Y(t_f)+Y(t_i)   \right]_{\sigma = \pi} - 2 \pi 
mR
\end{equation}
which is the condition that $\chi$ is a map from the world sheet to $S^1$ with 
radius R.  The construction yields
\begin{equation}	
R^\prime =\frac{\alpha^\prime}{R} \: ,
\end{equation}

The global restriction is implemented here through the boundary conditions.
We have shown that the dual action to the open string theory requires an additional 
topological term in the action in order to obtain the correct boundary condition.

The construction of global duality maps required then the implementation of a global
constraint which in the case of S-duality ensures the existence of local p-forms with
nontrivial transitions on a higher order bundle.  In the case of T duality the global
constraint is related to the compactification of one or several of the target
coordinates.  In the duality equivalence of the d=11 supermembrane and the 
d=10 IIA
Dirichlet supermembrane the global constraint for the d=11 supermembrane is the
compactification condition while the global constraint for the Dirichlet supermembrane
ensures that the local 1-form A is a connection on a nontrivial line bundle over the
world-volume.
In the construction of duality maps between p-forms and (d-p-2)-forms the difficult but
crucial step in the construction is the converse theorem that ensures that given a
globally defined (p+1)-form $L_{p+1}$ there exists a bundle of an order p and a local
antisymmetric field with non-trivial transitions whose curvature is $L_{p+1}$.  In the case
of p=2 there is a very elegant construction of the higher order bundle in terms of
Dixmier-Douady sheaves of groupoids \cite {5}.

The main result of the global analysis we have considered is that the existence of a
quantum equivalent dual theory completely determines the configuration space of the
potentials $A_p$  and of its local dual forms $V_{d-p-2}$.
The global constraint we have introduced are just the correct ones to describe the global
structure of the configuration spaces.   The geometrical description of these spaces
allow an explicit formulation of the D-brane theories in terms of the potentials $A_p$, a
necessary step for the quantization of the these theories.
\section{Non abelian duality}

Duality maps between abelian gauge theories given by $U(1)$ connections on line bundles over a
manifold $X$ can be shown to exist  by using a  quantum equivalent formulation of the original
theory in terms of closed 2-forms. This is expressed as a functional on the space of abelian
2-forms which must be  constrained by non-local restrictions, namely, the requeriment of being
closed and with integral periods, ensuring the existence of a 1-1 correspondence between the
space of constrained 2-forms and the line bundles over $X$ \cite{3}.This procedure has
been successfully applied even to more general U(1) bundles \cite{2} based on an
extension of Weil's theorem to complex p-forms \cite{Bry}.Once the equivalence between the
formulation in terms of the configuration space of abelian connections and that of the space
of closed 2-forms is achieved, the latter is used to construct at the quantum level the dual
gauge theory, by introducing dual Hodge-${\star}$ forms through Lagrange multipliers proving
the existence of non trivial relations between the partition functions of the abelian theory
and its dual.

The purpose of this talk is to inquire on the possibility of extending the above procedure to
the non abelian case. In the first place, we will begin by asking what conditions should be
imposed on matrix-valued 2-forms over a manifold $X$ so that we could produce something
similar to Weil's theorem for non-abelian 2-forms, so that we could achieve an equivalence
between the formulation of the theory on the configuration space of connections and the
formulation on the space of 2-forms. This actually is a formidable problem still not solved
but only to the level of conjectures \cite{Bry2}. In any case, we could try to see where
failures lie. 

The Bianchi identity for a matrix-valued 2-form $\Omega $ \[ {\cal D}\;\Omega =
0 \] is the first condition that comes to mind when looking for restrictions to implement, since
curvatures for connections on fiber bundles satisfy it. But this, in general, does not assure even that $\Omega$ may be expressed
locally in terms of any 1-form connection $A$ as \[ \Omega = dA + A\wedge A \] something
equivalent to a Poincar\'e's lemma for "covariantly closed" forms does not hold. Moreover, even
when we could express $\Omega$ in terms of $A$ as above, on open sets $U_i$ of a covering of the manifold
$X$, compatibility of the curvature-like 2-forms $\Omega(A_i)$ and $\Omega(A_j)$ on the
intersection of two open sets $U_i$ and $U_j$ should imply that $A_i$ and $A_j$ are related by
a well defined gauge transformation on the intersection of open sets. Simple calculations show
that this is not the case. $A_i$ and $A_j$ could be related by some other more general
transformations that no doubt include the mentioned gauge transformations i.e. \[ \Omega (A_i)
= g^{-1} \Omega(A_j)\; g \;  \stackrel{not}{\Longrightarrow}\; A_i = g^{-1}A_j \;g + g^{-1}dg
\]

Obviously, we need more restrictive conditions to arrive to the necessary compatibility
glueing for constructing globally well defined non abelian vector bundles.

It is well known \cite{Poly}, that a formulation of non abelian gauge theories has a rather
simple expression on the space of loops as a trivial flat gauge theory. The main ingredient in
this formulation is the use of the holonomy associated to each class of  non abelian Lie
algebra valued connections on a vector bundle. The use of holonomies  is quite adequate since
its non local character as a geometrical object carries a lot more information about the
bundle than curvatures or connections. So, we should go to the loop space formulation and see
whether it is possible to write some conditions that could characterize the non abelian
bundles and look for a procedure to build the duality maps. In what follows, we suceed in
proving half the task, for a more detailed discussion see \cite{LIA}.

For our purpose, instead of using the space of closed curves \cite{Poly,Barre}, we will consider
a space of open curves ${\cal C}$ with fixed endpoints {\rm O}, {\rm P} over a compact manifold
$X$. This will allow the construction of  smoothly behaving mathematical objects like
functionals, variations of functionals, 1-form connection functionals and so on, on open
neighborhoods of the space of curves. Particularly, we avoid regularization problems in the
definition of the gauge "potential" on path space.

First, any functional over this space will be denoted $ \widetilde {\Phi} ({\cal
C}_{\;\rm{O,P}}) $ and a variation or increment of this functional due to a deformation on the
curve leaving the endpoints fixed is defined as \[ \widetilde {\Delta} \widetilde {\Phi}
({\cal C}_{\;\rm{O,P}}) \equiv \widetilde {\Phi} ({\cal C}_{\;\rm {O,P}}+\delta {\cal
C}_{\;\rm {O,P}}) - \widetilde {\Phi} ({\cal C}_{\;\rm{O,P}}) \] 
Deformations on the curves are smooth vector fields on open neighborhoods of $X$ where the curve ${\cal C}_{\;\rm{O,P}}$
lies,  tending to zero on the endpoints of the curve. We could relax this definition allowing
non zero deformations on one of the endpoints but then  we would need to impose a non linear
condition to get the compatibility requirement on the patching of the vector bundle
\cite{LIA}.
 Our version of holonomy is $H_A$, the path ordered exponential of a 1-form
connection $A$ over $X$ integrated over the open curve ${\cal C}$ i.e 
\[  H_{A} ({\rm O},{\rm P},{\cal
C}) \equiv exp :\int_{\rm O}^{\rm P}A : \]
it becomes the ordinary holonomy when {\rm O} and {\rm P} are identified.
$ \widetilde {\cal A}({\cal C}_{\;\rm{O,P}})$ denotes the 1-form connection functional acting on
deformations $S$. It is obtained from $H_A$
\[ \widetilde {\cal A}({\cal C}_{\;\rm{O,P}}) = - \widetilde {\Delta} H_A \cdot H_{A} ^{-1}
\]
and may be expressed in terms of $ F_{p(t')}$ , the ordinary pointwise defined curvature 2-form
associated to the connection $A$, as

\[ \widetilde {\cal A}({\cal C}_{\;\rm{O,P}}) [S]
= \int_{\rm O}^{\rm P} H_{A} ({\rm O},p(t'),{\cal C}) F_{p(t')} [T,S] H_{A} ({\rm
O},p(t'),{\cal C}^{-1}) dt' \] 
where $T$ is a vector field tangent to the curve ${\cal C}$, $t'$ is a parameter along the curve
and $ p(t')$ is an ordinary point on the curve.

$ \widetilde {\cal A}({\cal C})$ is defined for classes of equivalence of ordinary connections
under gauge transformations, i.e it is gauge invariant up to elements of the structure group on
the endpoints of the curve. This is a rather nice feature of working in 
path spaces.

We could continue and define also the curvature functional $ \widetilde {\cal F} ({\cal
C},A)$ for the connection functional $ \widetilde {\cal A}$ in the usual manner

\[ \widetilde {\cal F} ({\cal C},A) = \widetilde {\Delta} \widetilde {\cal A} ({\cal C})
 \; + \; \widetilde {\cal A} ({\cal C}) \wedge \widetilde {\cal A} ({\cal C})  \]
 
for this free formulation of non abelian gauge theories, calculations show that
\[\widetilde {\cal F} ({\cal C},A) \;=\; 0 \]
and it is a gauge invariant statement.

The "covariant" derivative $\widetilde{\cal D}$ may be also introduced as
\[ \widetilde{\cal D} \;\cdot \;\; \equiv \;\; \widetilde {\Delta}\;\cdot\;\; +\; \widetilde
{\cal A} \wedge\; \cdot \]

In the case of the space of curves with two fixed endpoints, we need only to require
compatibility of $ \widetilde {\cal A}({\cal C}_{\;\rm{O,P}})$ on the intersecting neigborhoods

\[ \widetilde {\cal A}({\cal C}_{\;\rm{O,P}})_i \;\;\; = \;\;\;\widetilde {\cal A}({\cal
C}_{\;\rm{O,P}})_j\]

and get, in the same manner, that this is only possible if and only if
\[A_i = g^{-1} A_j\;g + g^{-1}d\;g.\]

So we have suceeded in the first step towards the construction of dual non abelian fields
suggesting that the natural space for building up the dual maps are loop spaces or open curve 
spaces. Now, it rests to find a global condition equivalent to that of integral periods of the
curvature 2-form for abelian gauge theories, that actually labels the different line bundles,
i.e an equivalent Dirac quantization condition. We know that for particular $SU(2)$ bundles, 
there may be a splitting  into the direct sum of two line bundles, for those bundles the usual Dirac
quantization may suffice. In the space of paths then the restriction to be imposed would be that
the ordinary curvature 2-form appearing in $ \widetilde {\cal A}({\cal C}_{\;\rm{O,P}})$ would
belong to the set of "diagonalizable" 2-form curvatures through a condition involving the
intersection form and the non abelian topological charge associated to the second Chern class.
This suggests that perhaps the global condition needed for non abelian gauge theories, at least
for the case of $SU(N)$, involves the "quantization" of the topological charge  associated to
the second Chern class. Once we find the exact condition to be imposed on the 
path space, the
dual map for constructing dual gauge theories should be no problem since the operation that
generalizes the Hodge-$\star$ operation for path spaces has already been defined 
at least on shell in \cite{Cho} and improved in \cite{LIA} in the sense 
that no regularization is needed.
The partition function is also easily implemented in our formulation. A characterization of
matrix-valued 2-forms for being curvatures of non abelian bundles has also recently been
conjectured using partial differential equations on a loop space \cite{Bry2}.

\end{document}